\documentclass[10pt,twocolumn]{article}

\usepackage[utf8]{inputenc}
\usepackage[english]{babel}
\usepackage{amsmath}
\usepackage{amsfonts}
\usepackage{amssymb}
\usepackage{mathpazo} 
\usepackage[scaled]{helvet}
\usepackage[T1]{fontenc}
\usepackage{url}
\usepackage{verbatim}
\usepackage{float}
\usepackage{caption}
\usepackage{subcaption}
\usepackage{graphicx}
\usepackage{xcolor}
\usepackage{booktabs}
\usepackage{algorithm}
\usepackage[noend]{algpseudocode}
\usepackage{changepage}
\usepackage[normalem]{ulem}
\usepackage{soul}
\usepackage[left=48pt,%
        right=42pt,%
        top=42pt,%
        bottom=60pt,%
        headheight=15pt,%
        headsep=10pt,%
        letterpaper,twoside]{geometry}%
        
\usepackage[labelfont={bf,sf},%
        labelsep=period,%
        figurename=Fig.,%
        singlelinecheck=off,%
        justification=RaggedRight]{caption}
        
\usepackage[numbers,sort&compress]{natbib}
\title {Generation of mega-gauss axial and azimuthal magnetic fields in a solid plasma by ultrahigh intensity, circularly polarised femtosecond laser pulses} 

\usepackage{authblk}
\author[1]{Anandam Choudhary}
\author[2]{Laxman Prasad Goswami}
\author[1]{C. Aparajit}
\author[1]{Amit D. Lad}
\author[1]{Ameya Parab}
\author[1]{Yash M. Ved}
\author[2,*]{Amita Das}
\author[1,*]{G.Ravindra Kumar}

\affil[1]{Tata Institute of Fundamental Research, 1 Homi Bhabha Road, Colaba, Mumbai 400 005, India}
\affil[2]{Indian Institute of Technology, Delhi }
\affil[*]{Corresponding author:amita@iitd.ac.in; grk@tifr.res.in}

\date{}

\begin{document}

\twocolumn[
 \begin{@twocolumnfalse} 
  
  \maketitle
  
\begin{abstract}

  The interaction of intense linearly polarized (LP) femtosecond laser pulses with solids is known to generate azimuthal magnetic fields, while circularly polarized (CP) light has been shown to create axial fields. We demonstrate through experiments and particle-in-cell simulations that circularly polarized light can generate both axial and azimuthal fields of comparable magnitude in a plasma created in a solid. Angular distributions of the generated fast electrons at target front and rear show significant differences between the results for the two polarization states, with circular polarization enforcing more axial confinement. The measurement of the spatial distribution of both types of magnetic fields captures their turbulent evolution.

  \vspace{2mm}
\end{abstract}

 \end{@twocolumnfalse}
]

\section{Introduction}
 Some of the largest terrestrial magnetic fields are generated in intense laser-produced plasmas on solid targets \cite{tatarakis2002measuring, wagner2004laboratory, 2020-4-Amita-RMPP-summary}. These quasi-static fields provide us opportunities to test physics in high magnetic fields akin to that in intrastellar, interstellar, and intraplanetary environments \cite{kaw2017nonlinear, drake2010high, glenzer2017preface, HEDLA2024}. They can also provide us applications in controlling electron transport in fusion targets \cite{kolka1993inertial, wang2015magnetically, sakata2018magnetized} and create unusually large spin domains in magnetic materials \cite{sinha2008mapping, nath2021macroscopic}. The interest in the experimental exploration of magnetic fields in laser-produced plasmas dates back to the 1960s with the pioneering work of \cite{korobkin1966investigation} and \cite{stamper1971spontaneous} among others. 

\begin{figure*}
  \centering
  \includegraphics[width=\linewidth]{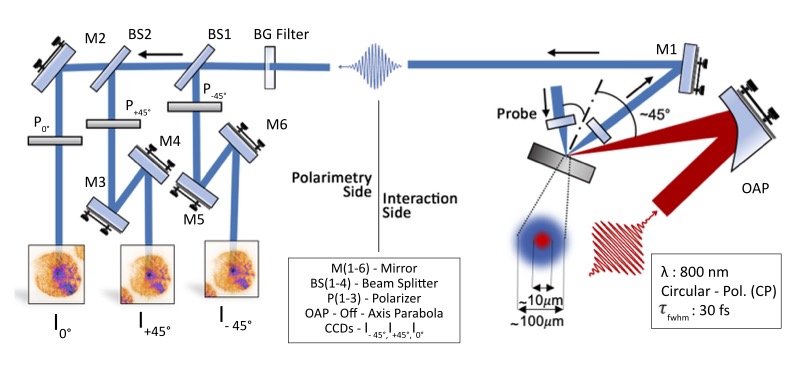}
  \caption{Experimental setup and process flow. The main pump laser irradiates the solid target (right side) and the second harmonic probe interacts with the plasma. The reflected probe goes to polarimetry setup and its polarization state is characterized using linear polarisers with different orientation angles. Pump and probe pulses are time and space synchronized.}
  \label{fig:exp_setup}
\end{figure*}

 Azimuthal magnetic fields caused by fast electron transport and the resulting instabilities, resonance absorption of laser light, and 
 the laser ponderomotive force has been explored in various models, simulations \cite{stamper1991review, haines2001generation,steiger1972intensity, zeng1999magnetic, romanov2004magnetic, abdullaev1986generation, chakraborty1988scaling, shvets2002magnetic, bhattacharyya1998spontaneous, kostyukov2001inverse, gorbunov1998magnetic, sudan1993mechanism} and some experiments \cite{sandhu2002laser, kahaly2009polarimetric, chatterjee2012macroscopic, mondal2012direct, chatterjee2017magnetic, das2020boundary, forestier2019space} in the last few decades both with nanosecond and femtosecond pulses, for their own sake as well as for simulating astrophysical phenomena in the laboratory. Femtosecond pulses are particularly important for the observation of the dynamics of the fast electrons generated in the plasma \cite{sandhu2002laser, sandhu2006real, chatterjee2012macroscopic, das2020boundary}. Fast electrons are expected to be important for fast ignition of laser fusion, hard x-ray, high energy ion, positron, and neutron sources \cite{rocca2024ultra}. Since the dynamics of the fast electrons manifest in and are controlled by their self-generated magnetic fields, it is important to map out the fields in space and time \cite{mondal2012direct, shaikh2016megagauss, chatterjee2017magnetic}.


On the other hand quasi-static axial fields are useful for efficient creation of ultra-high-energy-density states \cite{sakata2018magnetized} or confining relativistic electron beams \cite{cai2013effects, choudhary2023controlling}, laser propagation and energy absorption to bulk plasma regions \cite{goswami2021ponderomotive, goswami2022observations}, high harmonic generation \cite{dhalia2023harmonic, maity2021harmonic}, and magnetic field induced transparency \cite{mandal2021electromagnetic}. 

 Axial magnetic fields generated by circularly polarized (CP) laser radiation via the inverse Faraday effect have been investigated in many theoretical studies in underdense and overdense plasmas \cite{haines2001generation,steiger1972intensity, zeng1999magnetic, romanov2004magnetic, abdullaev1986generation, chakraborty1988scaling, shvets2002magnetic, bhattacharyya1998spontaneous, kostyukov2001inverse, gorbunov1998magnetic}. Experimental studies  with this polarization state are far fewer, many with nanosecond pulses on solid targets at normal laser  incidence\cite{horovitz1997measurements, eliezer1992generation} and a few with femtosecond pulses but in underdense plasmas \cite{najmudin2001measurements}. There  does not appear any study that measures both axial and azimuthal fields under the same excitation, and that too with circularly polarized, obliquely incident relativistic intensity femtosecond laser pulses in a solid plasma. We address this problem in this paper.  
 



\section{Experiment}

The experiment used the 150 TW laser system at the Tata Institute of Fundamental Research, Mumbai. A laser pulse at a central wavelength of 800 nm and duration of 30 femtoseconds was directed onto an Al-coated BK7 target at a $45^\circ$ angle of incidence, focused to a spot size of $7 \mu$m using an $f/3$ off-axis parabolic dielectric mirror. The peak intensity generated was $2.2 \times 10^{19}$ W/cm$^2$. The laser intensity contrast was $10^{-7}$ at 20 picoseconds before the peak of the pulse.  The schematic of the experimental setup is depicted in Figure \ref{fig:exp_setup}.

\begin{figure*}[!ht]
\centering
  \begin{subfigure}{.48\linewidth}
    \centering
		\includegraphics[width=\columnwidth]{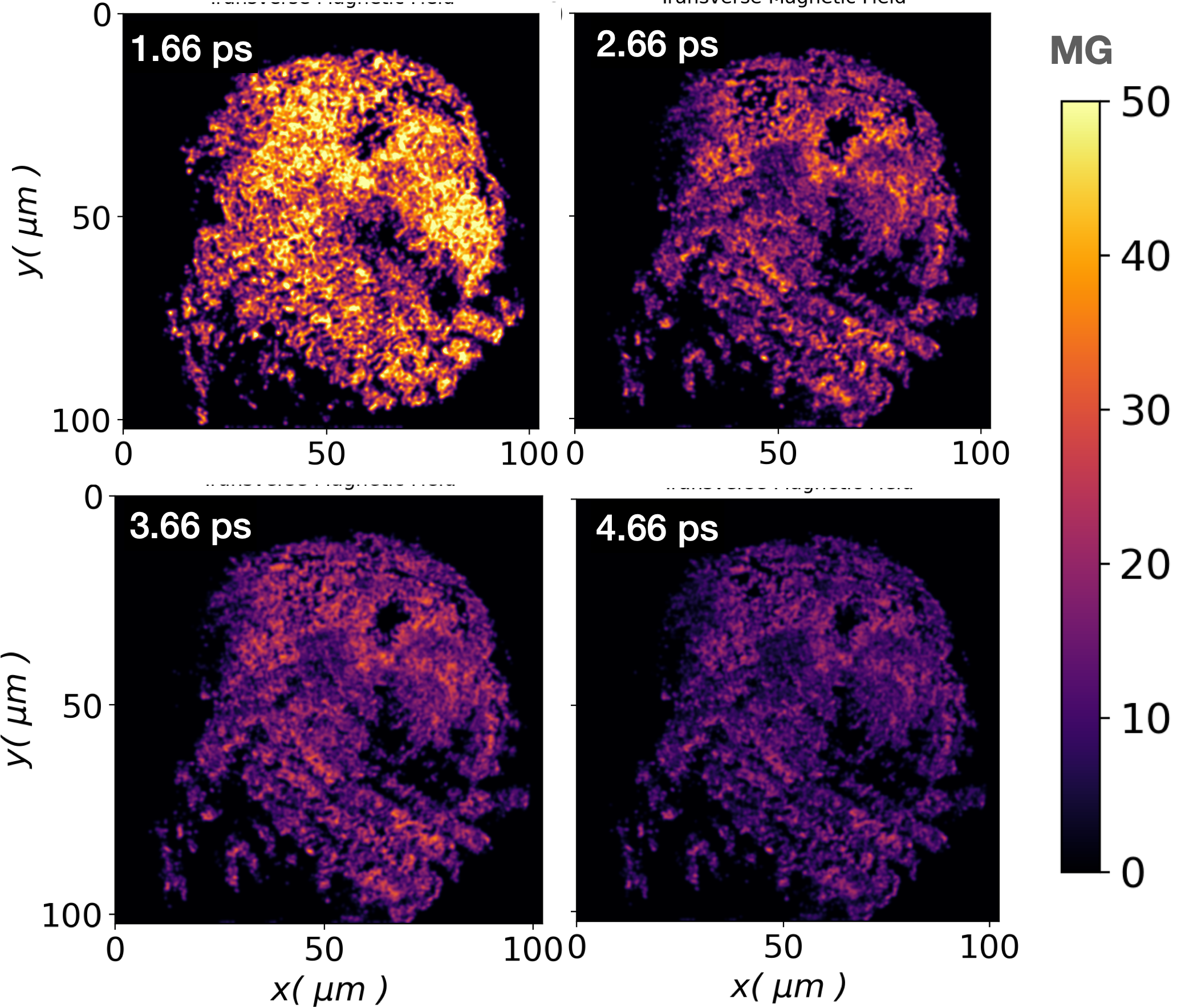}
		\caption{}
		\label{fig:B_trans2d}
  \end{subfigure}
  \begin{subfigure}{.48\linewidth}
    \centering
    \includegraphics[width =\columnwidth]{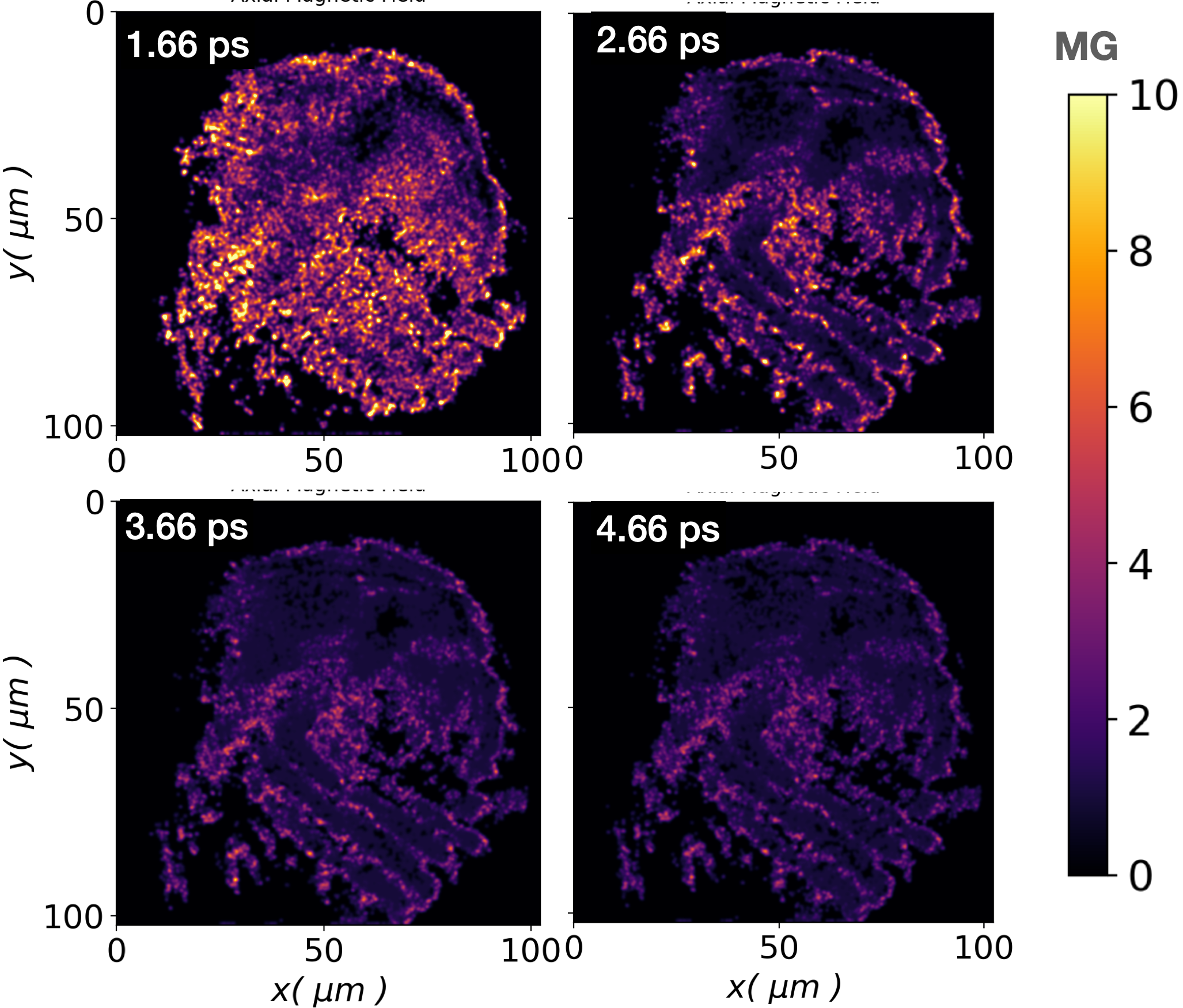}
    \caption{}
    \label{fig:B_axial2d}
  \end{subfigure}
  \caption{(a) spatial profile of transverse magnetic field with the circularly polarized laser pulse. (b) Axial magnetic field spatial profile for circularly polarized laser pulse generated plasma. Fields are in units of Mega Gauss.}
\end{figure*}

The probe pulse is derived from the primary interaction pump through a thin beam-splitter  and its frequency is doubled (refer to Figure \ref{fig:exp_setup} and attenuated to a lower intensity (less than $10^{11}$ W/cm$^{2}$). It is converted  to the second harmonic using a Type I $\beta$-barium borate (BBO) crystal. The probe is temporally delayed with respect to the pump by a retro-reflector mirror mounted on a controlled delay stage (Physik Instruments). Our temporal measurement resolution is confined to the width of the probe pulse ($\sim$ 200 fs). Additionally, the second harmonic probe can penetrate the overdense plasma up to 4$n_c$ (where $n_c$ represents the critical density of the primary interaction pulse). The probe is focused  to a spot of $80\mu$m size  at near normal incidence on the target, aligning with the direction of the plasma density gradient. The pump and probe's spatial overlap and temporal synchronization were carefully monitored using visual aids.  A sharp transition in the reflectivity and transmissivity of the probe from the laser-created plasma determines the temporal concurrence of the pump and probe pulses. The polarization of the pump and probe pulses was controlled using appropriate waveplates and polarizers. The experiment was conducted inside a vacuum chamber at a pressure of $10^{-5}$ Torr. \smallskip

To measure the distribution of electron fluxes, image plates (IP) were positioned in a cylindrical configuration surrounding the central target. Each image plate was shielded with 10 layers of $11\mu$m thick aluminum foil to block low-energy electrons (up to 100 keV) and electromagnetic emissions in and around the visible region. The image plates captured electron flux counts, which can be analyzed using image scanners to retrieve the data \cite{tanaka2005calibration}.

\subsection*{Polarimetry and Data analysis}

The complete polarization state of the reflected probe is determined using the experimentally measured Stokes' vectors. When a laser-produced plasma is subjected to a transient magnetic field, it behaves as a birefringent medium, causing the probe pulse to split into ordinary and extraordinary modes. These modes acquire different phases, resulting in the reflected probe attaining general polarization states with finite ellipticity (Cotton-Mouton effect) and orientation angle rotation (Faraday effect) in the presence of transverse and axial magnetic fields, respectively. 
\smallskip

The schematic of the Stokes' vector measurement setup is illustrated in Figure \ref{fig:exp_setup}. The reflected probe image is appropriately magnified and divided into three parts using beam splitters and analyzed using a high-optical-quality Glan-air polarizer (Leysop, extinction ratio of $10^{-6}$) with three different angle orientations ($0^\circ$, $45^\circ$ and $-45^\circ$). The signals ($I_{45^\circ}$, $I_{-45^\circ}$ and $I_{0^\circ}$) on the detector (Andor iKon-M CCDs, 1024 x 1024 pixels, pixel size 13.5 $\mu$m) are obtained in terms of Stokes vectors in the following order: $I_{0^\circ} = \frac{1}{2}(S_0 + S_1)$, $I_{45^\circ} = \frac{1}{2}(S_0+S_2)$, and $I_{-45^\circ} = \frac{1}{2}(S_0 -S_2)$. Stokes components are related by  $S_0^2 = S_1^2+S_2^2+S_3^2$ where $S_0$, $S_1$, $S_2$, and $S_3$ are the Stokes vector components. 
\smallskip

Four unknowns are solved using four equations to obtain all four Stokes' components. The ellipticity ($\chi$) and Faraday rotation ($\psi$) values can be directly deduced from the measured Stokes' vectors using the formulas:
$$\chi = \frac{1}{2}tan^{-1}\left(\frac{s_3}{\sqrt{s_1^2+s_2^2}}\right)$$ 
$$\psi = \frac{1}{2}tan^{-1}\left( \frac{s_2}{s_1}\right)$$

Here $s_1$, $s_2$, and $s_3$ are the reduced Stokes' vector components (Stokes vector normalized with $S_0$). This polarization data was captured for different delays. \\

\textbf{Estimation of magnetic fields:} As the probe travels through the plasma up to the critical layer for 400 nm wavelength, it encounters both axial and transverse fields leading to changes in the Stokes' vector. Numerical integration of the Stokes' vector evolution equation \cite{segre1999review} over the plasma scale length provides a mapping between magnetic fields and changes in the Stokes' vector. The plasma scale length or plasma slab length ($L=c_st$; $c_s$ ion acoustic speed; $t$ probe delay with pump) is calculated for each probe delay. The ion acoustic speed ($c_s$) is estimated from separate measurements of Doppler shifts \cite{adak2014ultrafast, jana2018probing, shaikh2018tracking}. This mapping enables the determination of the best-fit values of magnetic fields corresponding to the experimentally measured Stokes' vector changes. Both transverse and axial magnetic fields are measured over the transverse span of the probe and each measurement is integrated over the plasma scale length. Detailed discussions of this technique are available elsewhere \cite{segre1999review}.

\begin{figure}[!ht]
  \centering
  \includegraphics[width=0.8\linewidth]{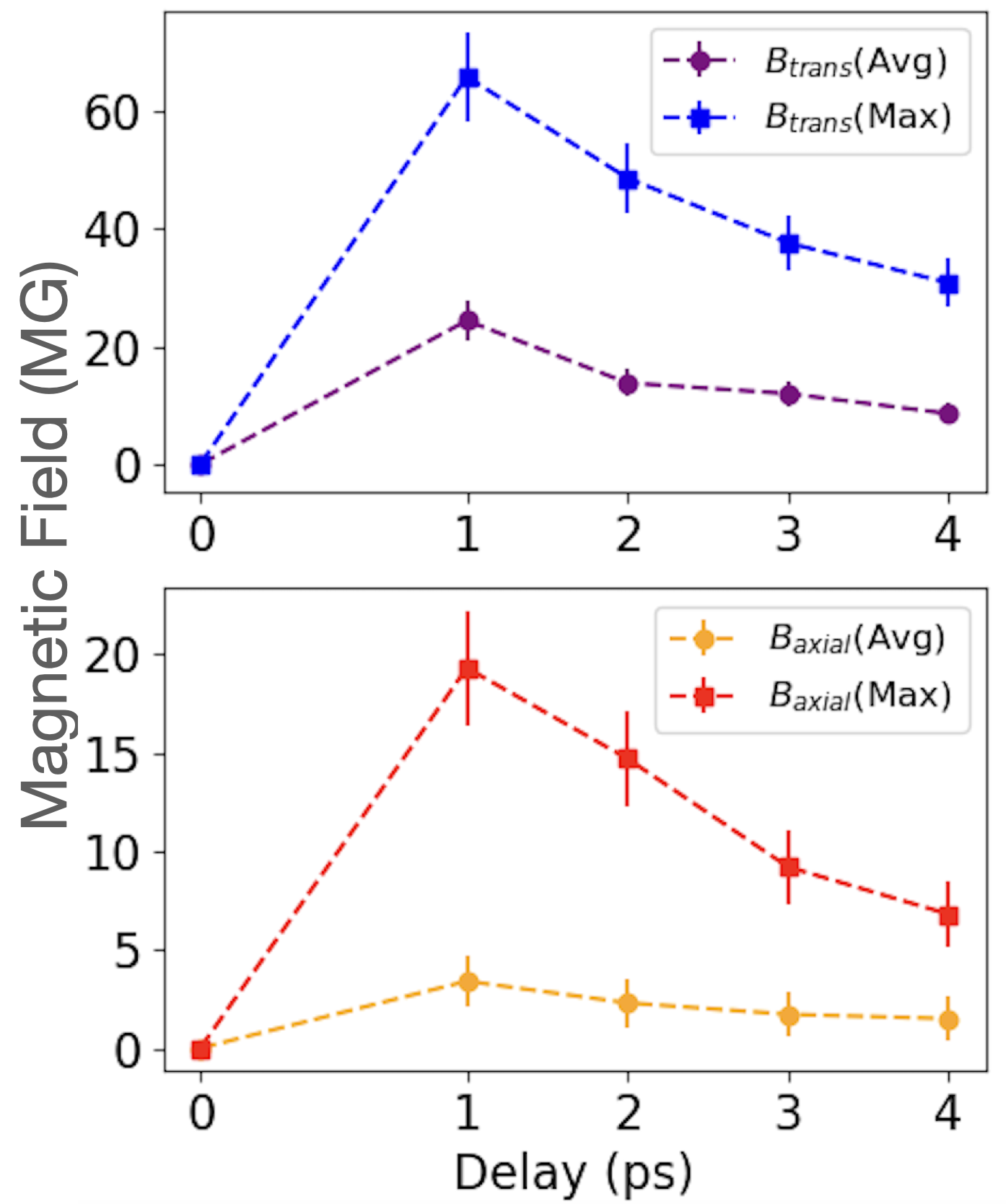}
  \caption{Integrated axial and transverse field for different delays with pump laser pulse.}
  \label{fig:Integrated_field}
\end{figure}

\begin{figure}[!ht]
  \centering
  \includegraphics[width=\linewidth]{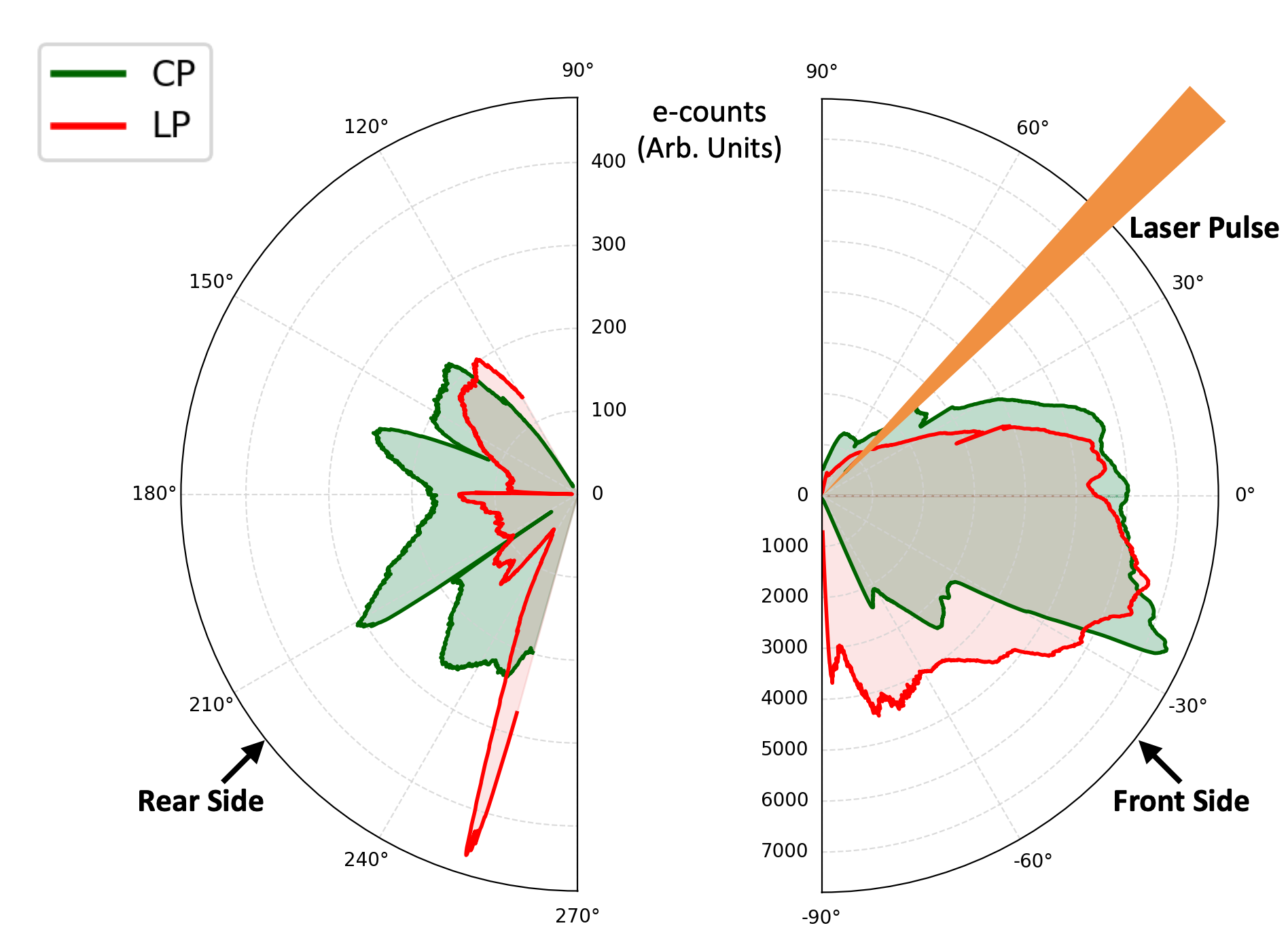}
  \caption{Electron angular distributions from linearly and circularly polarised laser pump irradiated targets at 45 degrees AOI.}
  \label{fig:AngDistr_experiment}
\end{figure}

\begin{figure}[!ht]
  \centering
  \includegraphics[width=0.9\linewidth]{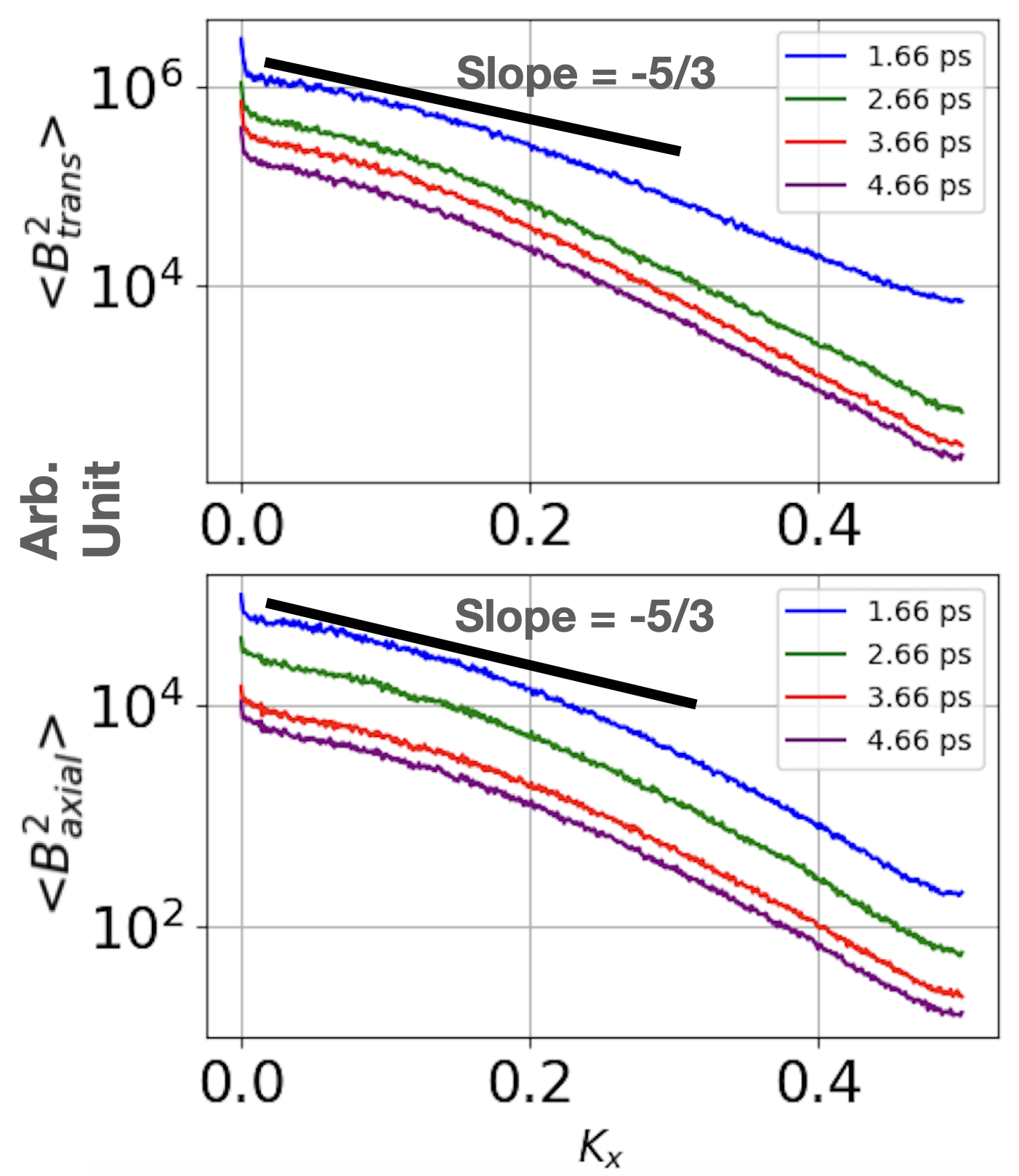}
  \caption{Magnetic field power spectrums for both axial and transverse fields at different delays. Black lines are reference spectrum with -5/3 power index(Kolmogorov's power law). < > shows averaged value along y-axis.}
  \label{fig:Power_spectrum}
\end{figure}

\section{Experimental observations}

The experimental figures displaying both the axial and transverse magnetic fields are depicted in Figure \ref{fig:B_trans2d} and \ref{fig:B_axial2d}, with a spatial resolution on the order of micrometers. These field mappings extend up to $4.7$ picoseconds. The spatial distribution of axial and transverse fields shows  randomness and inhomogeneity. These profiles provide a strong direct indication of turbulence  as discussed later \cite{chatterjee2017magnetic, mondal2012direct}. Averaged and max values of fields are also shown in Figure \ref{fig:Integrated_field} over the transverse span of the probe. It can be observed that transverse magnetic fields peak at $70 MG$ and decay over time, which is consistent with previous findings \cite{shaikh2016megagauss, mondal2012direct, kahaly2009polarimetric}.
In the case of axial magnetic fields, we observe magnitudes up to $20 MG$ followed by a decay, which is indicative of azimuthal currents generated via angular momentum transfer from CP laser pulses to electrons. Notably, other investigations with LP laser pulses \cite{shaikh2016megagauss, forestier2019space} did not reveal any axial magnetic field generation, while the transverse fields were observed.

We also conducted measurements of electron flux angular distributions upon laser pulse interaction with solid targets. Figure \ref{fig:AngDistr_experiment} illustrates these distributions for both LP and CP lasers.  We can see that electron flow along the surface direction is less with CP laser excitation  than that with LP laser excitation and this is attributed to the presence of axial magnetic fields.

We now study the spectral behavior of the field energy in the axial and transverse magnetic fields. The power spectra have been evaluated from the polarigrams obtained at picosecond time scales. The transverse profiles of magnetic fields (axial and transverse) are converted into spatial Fourier transform $B(k_x,k_y)$ 2D images. These images become power spectrum $P(k_x,k_y)$ images when B field values are squared at each $k_x$ and $k_y$.  1D power spectrums which are shown in Figure \ref{fig:Power_spectrum} are obtained as follows: 
$$Q(k_x)=\int P(k_x,k_y)dk_y$$

The spatial randomness of the  field and the significantly broad power spectra  indicate turbulence (Figure \ref{fig:Power_spectrum}). The spectral scaling has a power law decay close to Kolmogorov's $-5/3$ scaling (black solid lines superposed as reference). This suggests that both axial and transverse magnetic fields have a power spectrum showing magnetohydrodynamic  turbulence. Such spectra of magnetic fields have been observed earlier and have parallels with those from astrophysical systems (e.g. \cite{mondal2012direct, chatterjee2017magnetic}) 

 \section{Simulations}
 \begin{figure}[!ht]
 \centering
  \includegraphics[width=\linewidth]{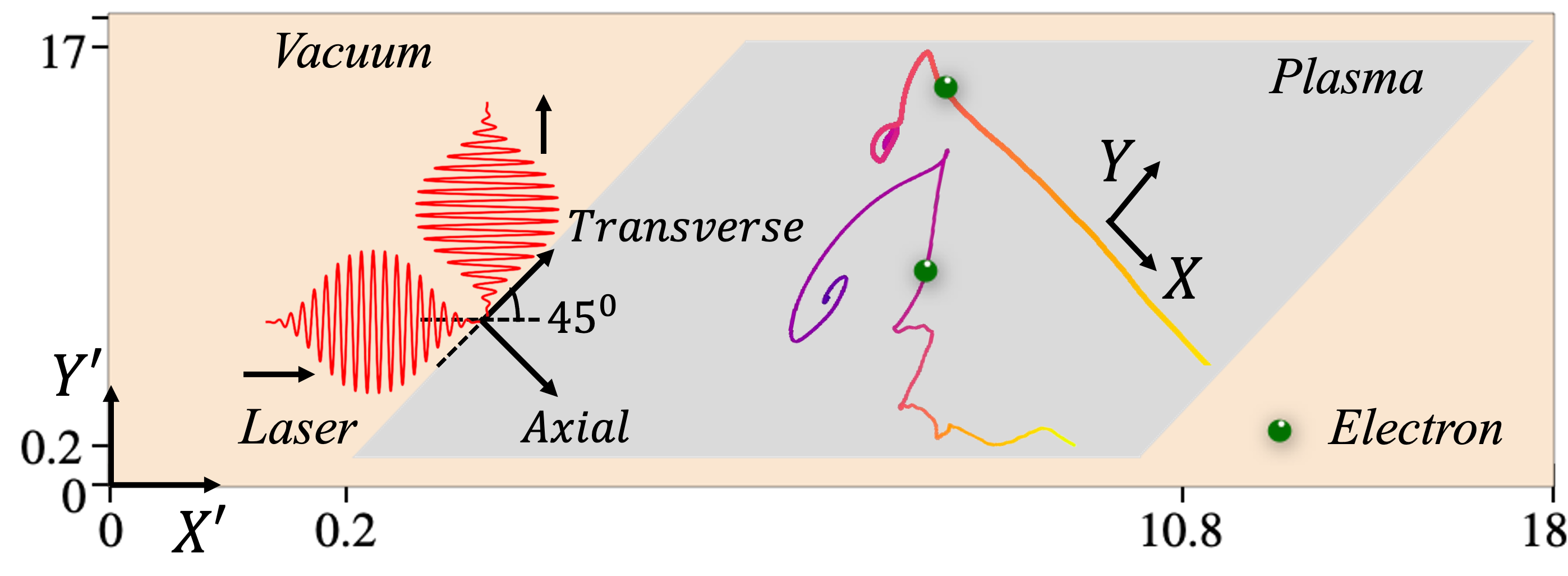}
 \caption{Figure shows the simulation geometry's schematics (not to scale). Here, the laser is incident at $45^0$ to the plasma surface. $X'$ and $Y'$ represent the laser propagation direction and its electric field. The $X$ and $Y$ are the target normal (therefore axial) and transverse directions. Two typical electron trajectories (one with LP and another with CP laser) have also been displayed.}
 \label{fig:SimulationSchematics}
\end{figure}

\begin{figure*}[!ht]
  \centering
  \begin{minipage}[c]{0.8\textwidth}
     \includegraphics[width=5.5in]{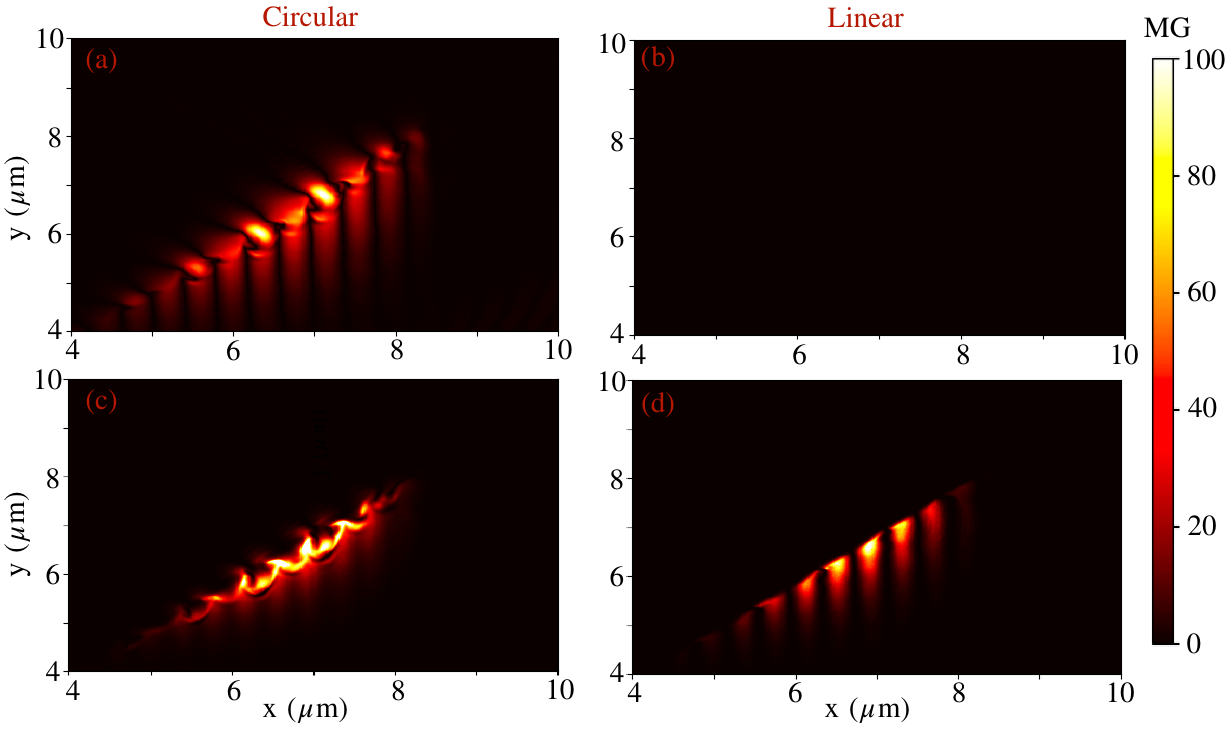}
  \end{minipage}\hfill
  \begin{minipage}[c]{0.20\textwidth}
     \caption{Figure shows the simulation results for magnetic field generation along the axial and transverse directions with the CP laser (in subplots a and c) and LP laser (in subplots b and d) at time $66fs$. Subplots (a) and (b) show the axial magnetic field, and subplots (c) and (d) show the transverse magnetic field.}
     \label{fig:MagneticAndCurrent}
  \end{minipage}
\end{figure*}

\renewcommand{\arraystretch}{1.3}
\begin{table}[!h]
	\caption{Values of various laser and plasma parameters}
	\label{table:simulationtable}
	\begin{center}
		\begin{tabular}{|c|c|c|}
			\hline		
			\color{red}Parameters & \color{red}Simulation & \color{red}Experimental\\
			\hline
			\multicolumn{3}{|c|}{\color{blue}Plasma Parameters} \\
			\hline
			$n_0$ & $1.0$ & $8.73\times10^{22} cm^{-3}$ \\
			$\omega_{pe}$ & $1.0$ & $1.67\times10^{16} rad/s$\\
			\hline
			\multicolumn{3}{|c|}{\color{blue}Laser Parameters} \\
			\hline
			$\omega_{l}$ & $0.14$ & $2.34 \times10^{15} rad/s$\\
			$\lambda_{l}$ & $44.7$ & $800 nm$\\
			Intensity & $a_0 = 4.0$ &$2.17\times 10^{19} W/cm^2$\\
			\hline
		\end{tabular}	
	\end{center}
\end{table}

While the experimental observations are over picosecond time scales, particle-in-cell (PIC) simulations can provide the dynamical picture at much shorter time scales and even while the femtosecond laser interacts with the plasma. With this objective, we  performed PIC simulations using the OSIRIS-4.0 platform \cite{hemker2000particle, fonseca2002osiris, fonseca2008one}, using both LP and CP laser pulses at $\lambda$ =800 nm incident at an angle of $45^\circ$ on the plasma surface from the left boundary (Figure \ref{fig:SimulationSchematics}). Since the OSIRIS PIC code does not permit oblique propagation in the simulation box, the target plasma surface has been tilted, as shown in the schematic Figure \ref{fig:SimulationSchematics}. The axial and transverse directions refer to the normal and the parallel to the plasma surface, respectively, as indicated by the $X$ and $Y$ directions in Figure \ref{fig:SimulationSchematics}. The tilted axis of $ X^{\prime} $ and 
$ Y^{\prime}$ directions refer to the laser propagation direction and the electric field direction. Since we intend to explore and understand the qualitative features of experimental observations using these simulations, the laser pulse duration comprises only $5$ laser cycles to 
optimize the computational resources at hand. The transverse spot size of $2.8\lambda (2.25 \mu m)$ is also smaller than the experimental value, while the normalized laser intensity ($a_0(=eE/m_e\omega_lc) = 4.0$) is used as per the experimental laser intensity of $2.17\times10^{19} W/cm^2$. We consider a fully ionized overdense plasma target (density $50n_c$) consisting of electrons and neutralizing heavy ions. The ions are held static in all the simulations. We have considered a simulation box size of  $22.5\lambda\times 22.5\lambda$. We introduce an exponentially sharp plasma density gradient profile of the form $n=50n_c(\exp((x-y)\ln(2)/L)-1)$ at the front plasma surface for a density scale length of $L=0.27\lambda (216 nm)$. The simulation parameters for the plasma and laser are summarized in table-\ref{table:simulationtable}.

\begin{figure}[!ht]
 \centering
  \includegraphics[width=0.8\linewidth]{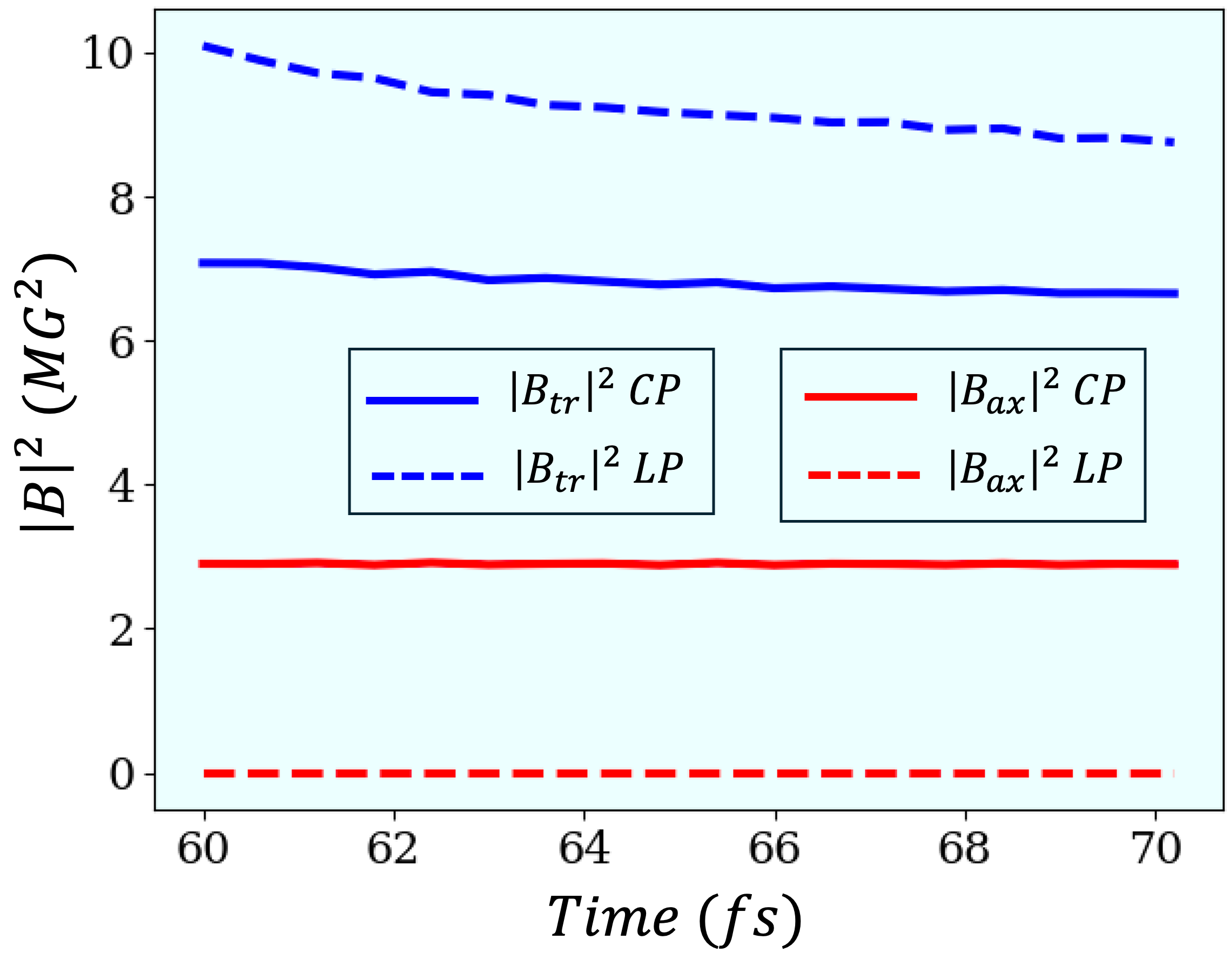}
 \caption{Figure shows the time evolution of the square of the axial and transverse magnetic field with the circular and linearly polarized laser pulse. Here, the initial energy of the LP and CP lasers is kept the same.}
 \label{fig:MagneticFieldEnerg}
\end{figure}

\begin{figure*}[!ht]
  \centering
  \includegraphics[width=5.0in]{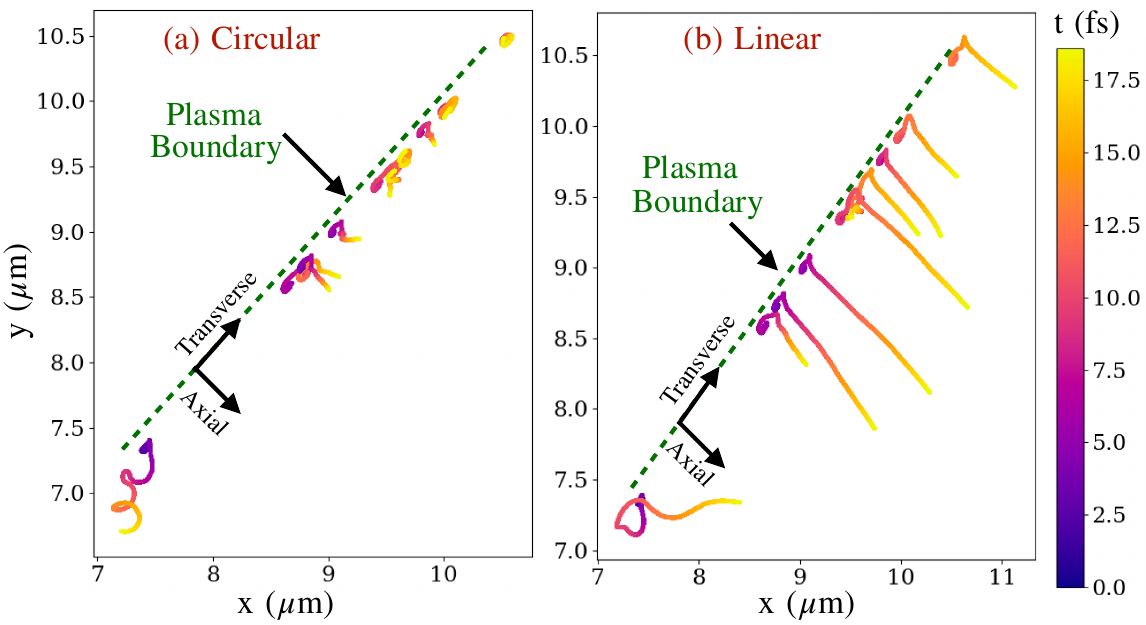}
  \caption{Figure shows the electron trajectories in XY-plane as a function of time (shown in color-bar) with (a) circularly and (b) linearly polarized laser pulse.}
  \label{fig:Tracking}
\end{figure*}

\begin{figure*}[!ht]
  \centering
  \includegraphics[width=6.0in]{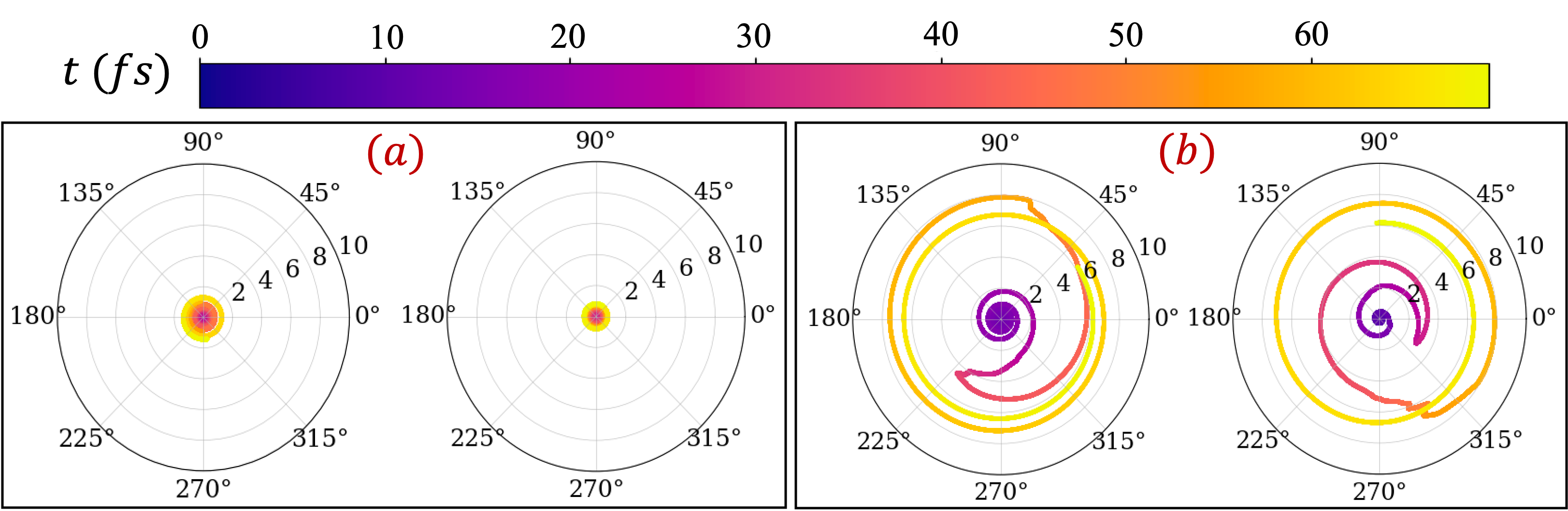}
  \caption{Figure shows the electron trajectories in $r-\theta$ plane as a function of time (shown in color-bar) with (a) circularly and (b) linearly polarized laser pulse.}
  \label{fig:PolarTracking}
\end{figure*}

Simulation results shown in Figure \ref{fig:MagneticAndCurrent} also corroborate the experimental findings of substantial generation of both transverse and axial magnetic fields with CP laser-coupled plasma (see subplots (a) and (c)). However, when the LP laser pulse is incident, as shown in subplots (b) and (d), there is no generation of the axial magnetic field (subplot(b) of the figure), though a transverse magnetic field can be observed (subplot(d)). The simulation results are presented at $66$ femtoseconds after the laser pulse has departed from the simulation box. The evolution of energy content in the transverse and axial magnetic fields for the CP and LP lasers has been depicted in Figure \ref{fig:MagneticFieldEnerg}. There is no energy content in the axial magnetic field for the case when the LP laser is interacting with the plasma target.


We now illustrate the behavior of electron trajectories originating from the target plasma surface when irradiated by the laser pulse. This has been done by considering a random sample of electrons emanating from the target surface. The trajectories of randomly selected electrons with both CP and LP laser polarizations are shown in Figure \ref{fig:Tracking}. In this figure, we have only shown the electron motion during the interaction of plasma with the laser pulse. A significant difference in electron trajectories can be observed when irradiated with CP and LP laser fields. The electron motion is predominantly axial for the LP laser, indicating a high axial speed (and hence axial current) leading to transverse magnetic field generation. On the other hand, for the CP laser, the transverse motion of electrons is quite dominant, which in turn would be responsible for the generation of the axial magnetic field as observed.  This also illustrates the evidence for angular momentum transfer from CP laser to electrons.

{Figure \ref{fig:PolarTracking} shows another perspective of these trajectories in a polar plot. The origin here corresponds to the initial location of each of the electrons. The radial excursion essentially shows the axial distance covered by the electrons from its initial location. The angular rotation with limited radial excursion in the case of CP laser suggests that the electron exhibits rotation in the transverse plane predominantly.}



\section{Conclusions}

We have demonstrated experimentally, the generation of both axial and transverse quasi-static, megagauss magnetic fields in a solid plasma created by circularly polarized, intense femtosecond laser pulses.   These measurements along with particle-in-cell simulations bring out the underlying electron dynamics in the two cases. Spatial and temporal mapping of the  fields indicates their turbulent evolution indicating possibilities for mimicking situations in other e.g. astrophysical contexts. \\

\noindent\begin{Large}
\textbf{Acknowledgement}\end{Large}
 GRK acknowledges partial support from the J.C. Bose Fellowship grant (JBR/2020/000039) from the Science and Engineering Board (SERB), Government of India. AD acknowledges support from the SERB core grants CRG 2018/000624 and CRG/2022/002782 as well as a J C Bose Fellowship grant JCB/2017/000055. AD and LPG would like to acknowledge the OSIRIS Consortium, consisting of UCLA and IST (Lisbon, Portugal) for providing access to the OSIRIS framework which is the work supported by NSF ACI-1339893. AD and LPG would like to thank the IIT Delhi HPC facility for providing the computational resources for simulation. LPG would like to thank the Council for Scientific and Industrial Research (Grant No-09/086/(1442)/2020-EMR-I) for funding the research. ADL acknowledges support from the Infosys-TIFR Leading Edge Research Grant (Cycle 2).




\bibliography{references}

\begin{thebibliography}{10}

\bibitem{tatarakis2002measuring}
M.~Tatarakis, I.~Watts, F.~Beg, E.~Clark, A.~Dangor, A.~Gopal, M.~Haines, P.~Norreys, U.~Wagner, M.-S. Wei, {\em et~al.}, ``Measuring huge magnetic fields,'' {\em Nature}, vol.~415, no.~6869, pp.~280--280, 2002.

\bibitem{wagner2004laboratory}
U.~Wagner, M.~Tatarakis, A.~Gopal, F.~Beg, E.~Clark, A.~Dangor, R.~Evans, M.~Haines, S.~Mangles, P.~Norreys, {\em et~al.}, ``Laboratory measurements of 0.7 gg magnetic fields generated during high-intensity laser interactions with dense plasmas,'' {\em Physical Review E}, vol.~70, no.~2, p.~026401, 2004.

\bibitem{2020-4-Amita-RMPP-summary}
{A.Das}, ``Laser plasma session:aapps-dpp conference, 12-17 nov 2018, kanazawa,'' {\em Reviews of Modern Plasma Physics}, vol.~4, Oct 2020.

\bibitem{kaw2017nonlinear}
P.~K. Kaw, ``Nonlinear laser--plasma interactions,'' {\em Reviews of Modern Plasma Physics}, vol.~1, pp.~1--42, 2017.

\bibitem{drake2010high}
R.~P. Drake, ``High-energy-density physics,'' {\em Physics Today}, vol.~63, no.~6, pp.~28--33, 2010.

\bibitem{glenzer2017preface}
S.~H. Glenzer, ``Preface to special topic: High-energy density laboratory astrophysics,'' {\em Physics of Plasmas}, vol.~24, no.~4, 2017.

\bibitem{HEDLA2024}
HEDLA2024, ``International conference on high energy density laboratory astrophysics (2022).'' \url{https://uia.org/s/ca/en/1300544400/}.

\bibitem{kolka1993inertial}
E.~Kolka, S.~Eliezer, and Y.~Paiss, ``Inertial plasma confinement in a miniature magnetic bottle induced by circularly polarized laser light,'' {\em Physics Letters A}, vol.~180, no.~1-2, pp.~132--136, 1993.

\bibitem{wang2015magnetically}
W.-M. Wang, P.~Gibbon, Z.-M. Sheng, and Y.-T. Li, ``Magnetically assisted fast ignition,'' {\em Physical review letters}, vol.~114, no.~1, p.~015001, 2015.

\bibitem{sakata2018magnetized}
S.~Sakata, S.~Lee, H.~Morita, T.~Johzaki, H.~Sawada, Y.~Iwasa, K.~Matsuo, K.~F.~F. Law, A.~Yao, M.~Hata, {\em et~al.}, ``Magnetized fast isochoric laser heating for efficient creation of ultra-high-energy-density states,'' {\em Nature communications}, vol.~9, no.~1, pp.~1--9, 2018.

\bibitem{sinha2008mapping}
J.~Sinha, S.~Mohan, S.~Banerjee, S.~Kahaly, and G.~R. Kumar, ``Mapping giant magnetic fields around dense solid plasmas by high-resolution magneto-optical microscopy,'' {\em Physical Review E}, vol.~77, no.~4, p.~046118, 2008.

\bibitem{nath2021macroscopic}
K.~Nath, P.~Mahato, A.~D. Lad, M.~Shaikh, K.~Jana, D.~Sarkar, R.~Sensarma, G.~R. Kumar, and S.~Banerjee, ``Macroscopic, layered onion shell like magnetic domain structure generated in yig films using ultrashort, megagauss magnetic pulses,'' {\em New Journal of Physics}, vol.~23, no.~8, p.~083027, 2021.

\bibitem{korobkin1966investigation}
V.~Korobkin and R.~Serov, ``Investigation of the magnetic field of a spark produced by focusing laser radiation,'' {\em Soviet Journal of Experimental and Theoretical Physics Letters}, vol.~4, p.~70, 1966.

\bibitem{stamper1971spontaneous}
J.~Stamper, K.~Papadopoulos, R.~Sudan, S.~Dean, E.~McLean, and J.~Dawson, ``Spontaneous magnetic fields in laser-produced plasmas,'' {\em Physical Review Letters}, vol.~26, no.~17, p.~1012, 1971.

\bibitem{stamper1991review}
J.~Stamper, ``Review on spontaneous magnetic fields in laser-produced plasmas: Phenomena and measurements,'' {\em Laser and Particle Beams}, vol.~9, no.~4, pp.~841--862, 1991.

\bibitem{haines2001generation}
M.~Haines, ``Generation of an axial magnetic field from photon spin,'' {\em Physical Review Letters}, vol.~87, no.~13, p.~135005, 2001.

\bibitem{steiger1972intensity}
A.~D. Steiger and C.~H. Woods, ``Intensity-dependent propagation characteristics of circularly polarized high-power laser radiation in a dense electron plasma,'' {\em Physical Review A}, vol.~5, no.~3, p.~1467, 1972.

\bibitem{zeng1999magnetic}
G.~Zeng and X.~He, ``Magnetic field generated by ionization front produced by intense laser radiating gas,'' {\em Physics of Plasmas}, vol.~6, no.~7, pp.~2954--2956, 1999.

\bibitem{romanov2004magnetic}
A.~Y. Romanov, V.~Silin, and S.~Uryupin, ``Magnetic field generation in a plasma produced through atomic ionization by circularly polarized radiation,'' {\em Journal of Experimental and Theoretical Physics}, vol.~99, pp.~727--732, 2004.

\bibitem{abdullaev1986generation}
A.~S. Abdullaev, I.~M. Aliev, and A.~Frolov, ``Generation of quasi-static magnetic fields by strong circularly polarized electromagnetic radiation in a relativistic magnetoactive plasma,'' {\em Fizika Plazmy}, vol.~12, pp.~827--835, 1986.

\bibitem{chakraborty1988scaling}
B.~Chakraborty, M.~Khan, B.~Bhattacharyya, S.~Deb, and H.~Pant, ``Scaling laws for a self-generated axial magnetic field in laser-produced plasma,'' {\em The Physics of fluids}, vol.~31, no.~5, pp.~1303--1305, 1988.

\bibitem{shvets2002magnetic}
G.~Shvets, N.~Fisch, and J.-M. Rax, ``Magnetic field generation through angular momentum exchange between circularly polarized radiation and charged particles,'' {\em Physical Review E}, vol.~65, no.~4, p.~046403, 2002.

\bibitem{bhattacharyya1998spontaneous}
B.~Bhattacharyya, P.~Mulser, and U.~Sanyal, ``Spontaneous faraday rotation due to strong laser radiation in a plasma,'' {\em Physics Letters A}, vol.~249, no.~4, pp.~324--329, 1998.

\bibitem{kostyukov2001inverse}
I.~Kostyukov, G.~Shvets, N.~Fisch, and J.~Rax, ``Inverse faraday effect in a relativistic laser channel,'' {\em Laser and Particle Beams}, vol.~19, no.~1, pp.~133--136, 2001.

\bibitem{gorbunov1998magnetic}
L.~Gorbunov and R.~Ramazashvili, ``Magnetic field generated in a plasma by a short, circularly polarized laser pulse,'' {\em Journal of Experimental and Theoretical Physics}, vol.~87, pp.~461--467, 1998.

\bibitem{sudan1993mechanism}
R.~Sudan, ``Mechanism for the generation of 10$^9$ g magnetic fields in the interaction of ultraintense short laser pulse with an overdense plasma target,'' {\em Physical review letters}, vol.~70, no.~20, p.~3075, 1993.

\bibitem{sandhu2002laser}
A.~Sandhu, A.~Dharmadhikari, P.~Rajeev, G.~R. Kumar, S.~Sengupta, A.~Das, and P.~Kaw, ``Laser-generated ultrashort multimegagauss magnetic pulses in plasmas,'' {\em Physical review letters}, vol.~89, no.~22, p.~225002, 2002.

\bibitem{kahaly2009polarimetric}
S.~Kahaly, S.~Mondal, G.~R. Kumar, S.~Sengupta, A.~Das, and P.~Kaw, ``Polarimetric detection of laser induced ultrashort magnetic pulses in overdense plasma,'' {\em Physics of plasmas}, vol.~16, no.~4, 2009.

\bibitem{chatterjee2012macroscopic}
G.~Chatterjee, P.~K. Singh, S.~Ahmed, A.~Robinson, A.~D. Lad, S.~Mondal, V.~Narayanan, I.~Srivastava, N.~Koratkar, J.~Pasley, {\em et~al.}, ``Macroscopic transport of mega-ampere electron currents in aligned carbon-nanotube arrays,'' {\em Physical review letters}, vol.~108, no.~23, p.~235005, 2012.

\bibitem{mondal2012direct}
S.~Mondal, V.~Narayanan, W.~J. Ding, A.~D. Lad, B.~Hao, S.~Ahmad, W.~M. Wang, Z.~M. Sheng, S.~Sengupta, P.~Kaw, {\em et~al.}, ``Direct observation of turbulent magnetic fields in hot, dense laser produced plasmas,'' {\em Proceedings of the National Academy of Sciences}, vol.~109, no.~21, pp.~8011--8015, 2012.

\bibitem{chatterjee2017magnetic}
G.~Chatterjee, K.~M. Schoeffler, P.~Kumar~Singh, A.~Adak, A.~D. Lad, S.~Sengupta, P.~Kaw, L.~O. Silva, A.~Das, and G.~R. Kumar, ``Magnetic turbulence in a table-top laser-plasma relevant to astrophysical scenarios,'' {\em Nature communications}, vol.~8, no.~1, p.~15970, 2017.

\bibitem{das2020boundary}
A.~Das, A.~Kumar, C.~Shukla, R.~K. Bera, D.~Verma, D.~Mandal, A.~Vashishta, B.~Patel, Y.~Hayashi, K.~Tanaka, {\em et~al.}, ``Boundary driven unconventional mechanism of macroscopic magnetic field generation in beam-plasma interaction,'' {\em Physical Review Research}, vol.~2, no.~3, p.~033405, 2020.

\bibitem{forestier2019space}
P.~Forestier-Colleoni, D.~Batani, F.~Burgy, D.~Del~Sorbo, F.~Froustey, S.~Hulin, E.~d'Humi{\`e}res, K.~Jakubowska, L.~Merzeau, K.~Mishchik, {\em et~al.}, ``Space and time resolved measurement of surface magnetic field in high intensity short pulse laser matter interactions,'' {\em Physics of Plasmas}, vol.~26, no.~7, 2019.

\bibitem{sandhu2006real}
A.~Sandhu, G.~R. Kumar, S.~Sengupta, A.~Das, and P.~Kaw, ``Real-time study of fast-electron transport inside dense hot plasmas,'' {\em Physical Review E}, vol.~73, no.~3, p.~036409, 2006.

\bibitem{rocca2024ultra}
J.~J. Rocca, M.~G. Capeluto, R.~C. Hollinger, S.~Wang, Y.~Wang, G.~R. Kumar, A.~D. Lad, A.~Pukhov, and V.~N. Shlyaptsev, ``Ultra-intense femtosecond laser interactions with aligned nanostructures,'' {\em Optica}, vol.~11, no.~3, pp.~437--453, 2024.

\bibitem{shaikh2016megagauss}
M.~Shaikh, A.~D. Lad, K.~Jana, D.~Sarkar, I.~Dey, and G.~R. Kumar, ``Megagauss magnetic fields in ultra-intense laser generated dense plasmas,'' {\em Plasma Physics and Controlled Fusion}, vol.~59, no.~1, p.~014007, 2016.

\bibitem{cai2013effects}
H.-b. Cai, S.-p. Zhu, and X.~He, ``Effects of the imposed magnetic field on the production and transport of relativistic electron beams,'' {\em Physics of Plasmas}, vol.~20, no.~7, 2013.

\bibitem{choudhary2023controlling}
A.~Choudhary, T.~Dhalia, C.~Aparajit, A.~D. Lad, A.~Dulat, Y.~M. Ved, R.~Juneja, A.~Das, and G.~R. Kumar, ``Controlling intense, ultrashort, laser-driven relativistic mega-ampere electron fluxes by a modest, static magnetic field,'' {\em arXiv preprint arXiv:2311.06884}, 2023.

\bibitem{goswami2021ponderomotive}
L.~P. Goswami, S.~Maity, D.~Mandal, A.~Vashistha, and A.~Das, ``Ponderomotive force driven mechanism for electrostatic wave excitation and energy absorption of electromagnetic waves in overdense magnetized plasma,'' {\em Plasma Physics and Controlled Fusion}, vol.~63, no.~11, p.~115003, 2021.

\bibitem{goswami2022observations}
L.~P. Goswami, T.~Dhalia, R.~Juneja, S.~Maity, S.~Das, and A.~Das, ``Observations of brillouin scattering process in particle-in-cell simulations for laser pulse interacting with magnetized overdense plasma,'' {\em Physica Scripta}, vol.~98, no.~1, p.~015602, 2022.

\bibitem{dhalia2023harmonic}
T.~Dhalia, R.~Juneja, L.~P. Goswami, S.~Maity, and A.~Das, ``Harmonic generation in magnetized plasma for electromagnetic wave propagating parallel to external magnetic field,'' {\em Journal of Physics D: Applied Physics}, 2023.

\bibitem{maity2021harmonic}
S.~Maity, D.~Mandal, A.~Vashistha, L.~P. Goswami, and A.~Das, ``Harmonic generation in the interaction of laser with a magnetized overdense plasma,'' {\em Journal of Plasma Physics}, vol.~87, no.~5, 2021.

\bibitem{mandal2021electromagnetic}
D.~Mandal, A.~Vashistha, and A.~Das, ``Electromagnetic wave transparency of x mode in strongly magnetized plasma,'' {\em Scientific Reports}, vol.~11, no.~1, pp.~1--11, 2021.

\bibitem{horovitz1997measurements}
Y.~Horovitz, S.~Eliezer, A.~Ludmirsky, Z.~Henis, E.~Moshe, R.~Shpitalnik, and B.~Arad, ``Measurements of inverse faraday effect and absorption of circularly polarized laser light in plasmas,'' {\em Physical review letters}, vol.~78, no.~9, p.~1707, 1997.

\bibitem{eliezer1992generation}
S.~Eliezer, Y.~Paiss, and H.~Strauss, ``Generation of a poloidal magnetic field by circularly polarized laser light,'' {\em Physics Letters A}, vol.~164, no.~5-6, pp.~416--418, 1992.

\bibitem{najmudin2001measurements}
Z.~Najmudin, M.~Tatarakis, A.~Pukhov, E.~Clark, R.~Clarke, A.~Dangor, J.~Faure, V.~Malka, D.~Neely, M.~Santala, {\em et~al.}, ``Measurements of the inverse faraday effect from relativistic laser interactions with an underdense plasma,'' {\em Physical Review Letters}, vol.~87, no.~21, p.~215004, 2001.

\bibitem{tanaka2005calibration}
K.~A. Tanaka, T.~Yabuuchi, T.~Sato, R.~Kodama, Y.~Kitagawa, T.~Takahashi, T.~Ikeda, Y.~Honda, and S.~Okuda, ``Calibration of imaging plate for high energy electron spectrometer,'' {\em Review of scientific instruments}, vol.~76, no.~1, 2005.

\bibitem{segre1999review}
S.~E. Segre, ``A review of plasma polarimetry-theory and methods,'' {\em Plasma physics and controlled fusion}, vol.~41, no.~2, p.~R57, 1999.

\bibitem{adak2014ultrafast}
A.~Adak, D.~R. Blackman, G.~Chatterjee, P.~Kumar~Singh, A.~D. Lad, P.~Brijesh, A.~Robinson, J.~Pasley, and G.~R. Kumar, ``Ultrafast dynamics of a near-solid-density layer in an intense femtosecond laser-excited plasma,'' {\em Physics of Plasmas}, vol.~21, no.~6, 2014.

\bibitem{jana2018probing}
K.~Jana, D.~R. Blackman, M.~Shaikh, A.~D. Lad, D.~Sarkar, I.~Dey, A.~P. Robinson, J.~Pasley, and G.~Ravindra~Kumar, ``Probing ultrafast dynamics of solid-density plasma generated by high-contrast intense laser pulses,'' {\em Physics of Plasmas}, vol.~25, no.~1, 2018.

\bibitem{shaikh2018tracking}
M.~Shaikh, K.~Jana, A.~D. Lad, I.~Dey, S.~L. Roy, D.~Sarkar, Y.~M. Ved, A.~P. Robinson, J.~Pasley, and G.~Ravindra~Kumar, ``Tracking ultrafast dynamics of intense shock generation and breakout at target rear,'' {\em Physics of Plasmas}, vol.~25, no.~11, 2018.

\bibitem{hemker2000particle}
R.~G. Hemker, {\em Particle-in-cell modeling of plasma-based accelerators in two and three dimensions}.
\newblock University of California, Los Angeles, 2000.

\bibitem{fonseca2002osiris}
R.~A. Fonseca, L.~O. Silva, F.~S. Tsung, V.~K. Decyk, W.~Lu, C.~Ren, W.~B. Mori, S.~Deng, S.~Lee, T.~Katsouleas, {\em et~al.}, ``Osiris: A three-dimensional, fully relativistic particle in cell code for modeling plasma based accelerators,'' in {\em International Conference on Computational Science}, pp.~342--351, Springer, 2002.

\bibitem{fonseca2008one}
R.~Fonseca, S.~Martins, L.~Silva, J.~Tonge, F.~Tsung, and W.~Mori, ``One-to-one direct modeling of experiments and astrophysical scenarios: pushing the envelope on kinetic plasma simulations,'' {\em Plasma Physics and Controlled Fusion}, vol.~50, no.~12, p.~124034, 2008.

\end{thebibliography}
\bibliographystyle{ieeetr}


\end{document}